\documentclass{article}

\usepackage[mathscr]{euscript}
\usepackage{amsmath}
\usepackage{color}
\usepackage[dvips]{graphicx}
\usepackage[top=1in, bottom=1.25in, left=0.75in, right=0.75in]{geometry}

\usepackage{subfig}

\newcommand{\red}[1]{\textcolor{black}{#1}}

\begin{document}
% \title{Stringer: Balancing Latency and Resource Usage in VNF Placement}
\title{Stringer: Balancing Latency and Resource Usage\\in Service Function Chain Provisioning}

\author{
    Freddy C. Chua, Julie Ward,\\
    Ying Zhang, Puneet Sharma, Bernardo A. Huberman\\
    Hewlett Packard Labs, Hewlett Packard Enterprise,\\
    Palo Alto, CA 94304, USA
}


\maketitle

\begin{abstract}
  Network Functions Virtualization (NFV) enables telecommunications infrastructure providers to replace special-purpose networking equipment with commodity servers running virtualized network functions (VNFs). A provider utilizing NFV faces the \red{Service Function Chain (SFC)} provisioning problem of assigning VNF instances to nodes in the physical infrastructure (e.g. datacenters), and routing Service Function Chains (sequences of functions required by customers, a.k.a. SFCs) in the physical network. The provider must balance competing goals of performance and resource usage. We present \red{an approach to SFC} provisioning, consisting of three elements. The first element is a \red{fast and scalable} round-robin heuristic. The second element is a Mixed Integer Programming (MIP) based approach. The third element is a queueing-theoretic model to estimate the average latency associated with any SFC provisioning solution. Our SFC provisioning system, called {\em Stringer}, allows providers to balance the conflicting goals of minimizing infrastructure resources and end-to-end latency for meeting their respective SLAs.
\end{abstract}


% \fi

% \IEEEpeerreviewmaketitle

\section{Introduction}
\label{sec:intro}

Telecommunications providers are making a strong push towards {N}{etwork} Functions Virtualization (NFV) of their infrastructure to reduce both CAPEX and OPEX while maintaining high carrier-grade service levels. The savings in CAPEX and OPEX come from being able to dynamically assign Virtualized Network Functions (VNFs) to various standard servers in their \red{infrastructure} to meet varying workload demands. Similar to Cloud Service resource allocation, such dynamic VNF placement can be automated to optimize various goals of a Telco Operator. Adoption of NFV by Telcos allows dynamic fine-grained Service Function Chaining (SFC) where various service functions chains can be strung together with deployed VNFs using SDN-enabled dynamic route control. \red{Similarly Enterprises are adopting NFV for deployment of network services in their infrastructure.}

% A few examples of VNFs are firewall services, intrusion detection, caches and proxies. In each SFC, traffic corresponding to the chain is dynamically steered through a sequence of VNFs.

% VNF comprising the particular SFC request

% We address the composite problem of orchestrating SFC provisioning within a datacenter that includes VNF placement and selection of VNF instances for each SFC that requires the respective VNFs.
\red{As a popular use case of NFV in the Telco, SFC is usually deployed in the Telco's datacenters in their PoPs or central offices. The prosperity of SFCs highly depend on its performance.} \red{We provide an approach to SFC provisioning within a datacenter. SFC provisioning comprises determining how VNFs are placed on nodes in the datacenter and how VNF instances are assigned to SFCs. The placement and assignment} affects the traffic routing from the SFC through the datacenter's network. Our SFC provisioning system, called {\em Stringer}, allows operators to balance the conflicting goals of minimizing infrastructure resources and end-to-end latency for meeting their respective SLAs.

In any SFC, the relative location of the VNFs will affect the end-to-end latency incurred by the packets traversing the particular SFC. A poor placement will cause the flow to traverse the same path-segments back and forth inside the network, increasing the network delay and consuming more bandwidth.
% Finding the best placement for all the VNFs and their respective instances in the context of service chaining is crucial to the carrier-grade performance of the entire NFV system.

% system flowchart
\begin{figure*}[htb]
  \centering
  \includegraphics[width=5.5in]{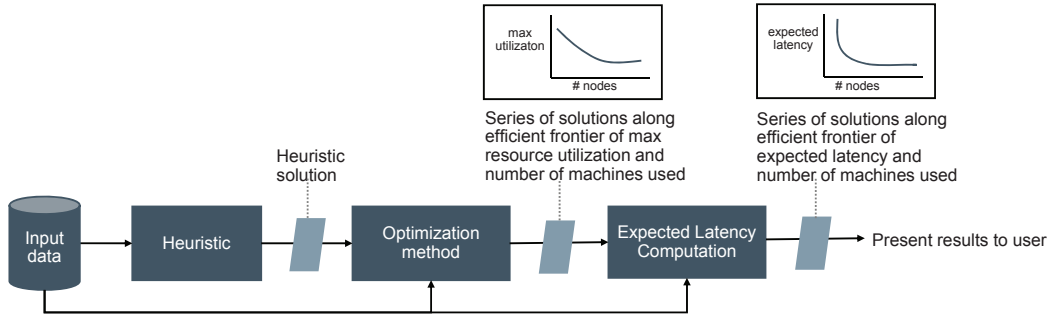}
  \caption{Flowchart of System: The input data, consisting of the service chains, their requirements and the data center network topology, are given to the heuristic. The basic solution from the heuristic provides an initial solution for the mixed integer program. Expected latency is then computed for each of the solutions from the optimizer. The user can then choose from a menu of solutions differing in expected latency and number of servers used.}
  \label{fig:system_flowchart}
\end{figure*}

The SFC provisioning problem entails choosing where to place instances of VNFs on servers in a NFV infrastructure to accommodate the traffic for a given set of SFC requests.
% Each service chain is a stream
\red{Each service chain is a sequence of VNFS that processes a stream of network packets flowing it at a certain rate} of network packets flowing through a sequence of VNFs at a certain rate. Network traffic for a given service chain must visit the chain's sequence of VNFs in the specified order. For example, a service chain may require packets to follow the VNF sequence: load balancer, network address translator, and firewall. In the SFC provisioning problem, one must place (possibly multiple) instances of each VNF on servers, and choose the route(s) for each service chain, in such a way that the network can accommodate the traffic for as many service chains according to their priorities.
\red{Service chains may share VNF instances. Moreover, the traffic for a given service chain may be split among multiple paths in the network when multiple instances of a specific VNF are used.}

\red{Our work differs from prior work in VNF placement in several important ways. One key difference is in the placement objective. Operators have multiple competing goals to consider when placing VNFs. A service provider may want to use as few servers as possible in order to minimize operating costs and leave open servers for future needs (\cite{Moens2014,Luizelli2015}). At the same time, the operator must ensure low \red{end-to-end} network latency for his customers.  These objectives are in direct conflict. While some prior work proposes multiple alternative objectives (\cite{Bari2015, Mehraghdam2014, COMB2012}), ours is the first, to our knowledge, that provides a flexible way to trade-off these competing goals in SFC provisioning. Moreover, unlike \cite{COMB2012}, our optimization model employs only linear constraints to model maximum utilization.}

\red{Another important difference is in the way packet delays are modeled. Most prior approaches \cite{Moens2014,Luizelli2015,Bari2015, Mehraghdam2014} model network latency with a known fixed delay when packets pass through VNFs, nodes and edges. They do not consider that latency depends on network traffic: packets traveling through congested network resources face much longer queueing delays than at uncongested ones. Expected latency depends on VNF placement and routing decisions, and the implied utilization of network resources, in a complex and non-linear way, which explains why prior work models latency in a simplified, utilization-independent way. Figure \ref{fig:rho_effects}, which shows the non-linear relationship between expected latency and utilization, highlights what is lost in this simplified approach. A single congested server or switch can dramatically increase latency for all service chains using that resource. If congestion is not explicitly modeled, such effects are ignored.}

\red{Some prior work (e.g., \cite{E22015}) decomposes VNF placement into two separate problems: first determining the number of instances of each VNF and routing among instances, and then placing VNF instances. Steering\cite{STEERING2013} assumes the number of instances of each VNF is given.  In contrast, our approach considers both problems simultaneously. Moreover, our approach considers the utilization of servers whereas \cite{E22015} considers only switch traffic.} \red{In this work, we focus on the chaining of inline services, e.g. firewall, load balancer, IDS. These VNFs operate on their own, with little dependencies across VNFs. Those VNFs with complex inter-dependences, e.g. EPC in cellular core network~\cite{rajan2015} are not the focus of this paper.}

\section{\em{Stringer}: Our SFC Placement System}
\label{sec:system}

% We describe various components of our system, called {\em Stringer}, for SFC provisioning.
{\em Stringer} provides the ability to the operators to select their operating point for trading-off resource usage and end-to-end SFC latency. Figure~\ref{fig:system_flowchart} shows the architecture and flowchart of {\em Stringer} system.
% It includes a mixed integer programming (MIP) model that explicitly captures the effect of network traffic on latency while maintaining a linear model: its objective minimizes the maximum utilization over resources in the network. Minimizing the worst-case utilization avoids the situation in which a small number of congested resources induce outsized delays on network traffic.

There are three main contributions of this work.  The first is a scalable heuristic that seeks to minimize the maximum utilization over all nodes. The second is our MIP-based placement approach that competing objectives of minimizing congestion-induced latency and minimizing the number of servers used. It minimizes a weighted combination of two metrics: (1) the number of servers used to host VNF instances, and (2) the maximum utilization over network resources, which we use as \red{a proxy for} latency.

The optimization method generates multiple SFC provisioning solutions for different relative weightings of the two objectives, thereby generating solutions along the efficient frontier of number of servers and maximum resource utilization. \red{The MIP and heuristic each have advantages: The MIP provides optimal benchmarks, but does not scale to very large size networks. The heuristic is fast and scalable, can be used to provide an initial solution that speeds up the MIP solution process, and generates solutions that are close to the efficient frontier of latency and node usage.}

%The main contribution of this work is a scalable and efficient placement algorithm and its evaluation. While the MIP can compute optimal solution, it cannot scale to a large size network. We overcome this challenge by having a two-step process. As illustrated in Figure \ref{fig:system_flowchart}, we first propose a fast, scalable round-robin heuristic for VNF placement, which generates an initial VNF placement solution. This initial result is then fed to the MIP to help speed up the search process. Secondly, our MIP algorithm balances the competing objectives of minimizing congestion-induced latency and minimizing the number of servers used. It minimizes a weighted combination of two metrics: (1) the number of servers used to host VNF instances, and (2) the maximum utilization over network resources, which we use as an approximation for latency. The optimization method generates multiple VNF placement solutions for different relative weightings of the two objectives, thereby generating solutions along the efficient frontier of number of servers and maximum resource utilization.

Our third contribution is a method to evaluate a SFC provisioning solution. Evaluating the performance of a placement strategy on a real world testbed of large size is not likely in practice. Therefore, we propose a queueing-theoretic model of the network which allows us to simulate the average expected latency associated with any given SFC provisioning solution under mild assumptions on the network traffic. Our model differs from standard M/M/1 queueing models in that it accounts for the fact that network elements have finite buffers and packets are dropped when they arrive to full buffers. Our expression for average expected latency reflects the possibility of packets being dropped and re-sent.

Combined, these three elements create our {\em Stringer} system that generates a set of SFC provisioning solutions varying in resource usage and performance. When presented with an array of solutions reflecting different tradeoffs between competing objectives, the operator can then make an informed choice about how to place VNFs and route the SFCs accordingly.

% computational results for a set of randomly generated problems.
\begin{figure*}[htb]
	\centering
	\includegraphics[width=5.0in]{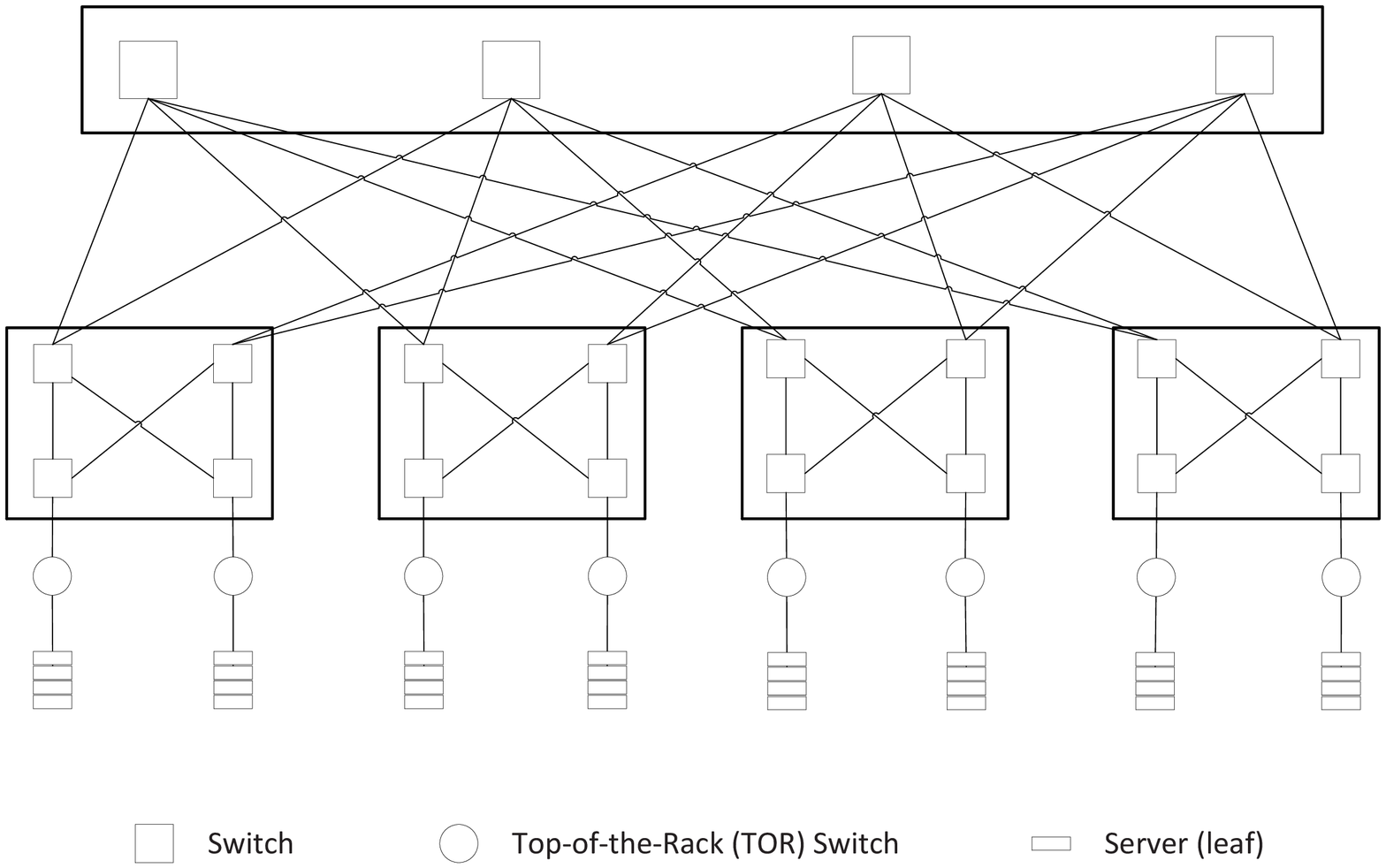}
	\caption[width=5.0in]{Tree Topology: The most common network topology used in datacenters is the FAT-tree topology where servers communicate with one another through a hierarchical arrangement of network switches as shown in the figure. The servers are connected to the TOR (top-of-the-rack) switches. Our model abstracts the underlying FAT tree topology by aggregating each level and cluster of switches into a single large virtual switch. This switch aggregation simplifies the model because each pair of servers has a unique path connecting them in the modified network.} %\cite{Al-Fares2008,Leiserson1985}
	\label{fig:topology}
\end{figure*}

\section{Preliminaries for Stringer}
\label{sec:prelim}
The inputs to Stringer fall into three categories:  the physical network topology, the virtualized network functions, and the service chains.
% temp

\paragraph*{Physical network} The physical network topology is a bi-directional graph with the property that each ordered pair of nodes has a unique acyclic directed path to each other. The underlying structure is a tree, consisting of switches (including a root switch $r$), and servers which are leaves in the tree.
% (Leaf nodes are nodes that are connected to only one other node in the network.)
Let $N$ denote the set of nodes (switches and servers) and $L \subset N$ be the set of servers. An example network is shown in Figure \ref{fig:topology}. Let $\mu_n$ be the processing rate, in packets per second, associated with any node $n \in N$.
% For every edge going from node $i$ to node $j$ in the network, there is a corresponding opposite edge going from $j$ to $i$.
 % Let $P_{n,m}$ be the set of nodes in the unique acyclic path from node $n$ to $m$, including the destination $m$ but excluding the origin $n$.
  %\setminus L$.
% Just as switches may have different processing rates, servers may also differ in their functionality. Let $S$ represent the set of different possible server types. Let $s_l$ be the server type associated with server $l \in L$.

\paragraph*{Virtualized Network Functions}  Let $V$ denote a set of VNF types. Instances of these VNF types must to be assigned to servers in the physical network in order to accommodate service chains. Multiple instances of a given VNF type $v$ may be assigned. We assume that a server in the network can accommodate at most one virtual network function instance, although that constraint can easily be relaxed.
% Performance of a given VNF depends on the machine type of the server to which it is assigned. To further simplify the analysis and optimization formulation, we assume that the VNFs do not add or drop packets for each service chain.

% Let $\gamma^s_v$ be the processing rate, in packets per second, of VNF type $v \in V$ when assigned to machine type $s \in S$.

\paragraph*{Service Chains} Let $C$ denote the set of service chains to be mapped to the network. Service chain $c \in C$ comprises a (possibly repeating) sequence of VNF types.
% Let $q_c$ represent the length (in number of VNFs) in service chain $c$. Let $\alpha^c_{i,v}$ be a binary parameter indicating whether the $i$th function in service chain $c$ is of type $v$.
The service chain $c$ is a Poisson process with arrival rate of $\lambda_c$ packets per second. Traffic for service chain $c$ enters the physical network through the root node, visits each function according to the chain's function sequence, and then departs the network from the root node. Let $\Lambda =  \sum_c \lambda_c$ be the sum of arrival rates of all service chains.

\section{Expected Latency Evaluation}
\label{sec:latency}
% \begin{figure*}[htb]
% 	\centering
% 	\includegraphics[width=6.0in]{eps/heuristic_flowchart.eps}
% 	\caption{Flowchart of the SFC provisioning Heuristic that maps the service chains to the underlying physical network. The entry point for this flow chart is at the rounded node labeled as ``Consider next service chain''. The flow chart will continue to loop infinitely until we run out of service chains to host.}
% 	\label{fig:heuristic_flowchart}
% \end{figure*}

We show how to compute the expected latency of any packet entering the system, assuming that VNF placement and service chain routing has already been determined.  The expected latency of a packet entering the network depends on the service chain with which the packet is associated. Let $E(T_c)$ represent the expected latency of packets in a given service chain $c \in C$. The expected latency $E(\mathscr{T})$ of a randomly selected arriving packet is equal to the sum over all service chains $c \in C$ of the probability $( {\lambda_c}/{\Lambda})$ that the packet is associated with chain $c$  times $E(T_c)$ :
\begin{align}
	E(\mathscr{T}) &= \sum_{c \in C} \frac {\lambda_c}{\Lambda} E(T_c) \\
	E(T_c) &:= E(T_{1 \rightarrow n})
\end{align}
where $E(T_{1 \rightarrow n})$ represents the expected latency for a packet to visit the sequence of nodes as $\{1,2,\ldots,n\}$ in $N_c$, for $n = 1,2,\ldots,|N_c|$.

\begin{figure}[htb]
	\centering
	\subfloat[Latency vs. utilization at node $n$.] {
		\includegraphics[width=2.6in]{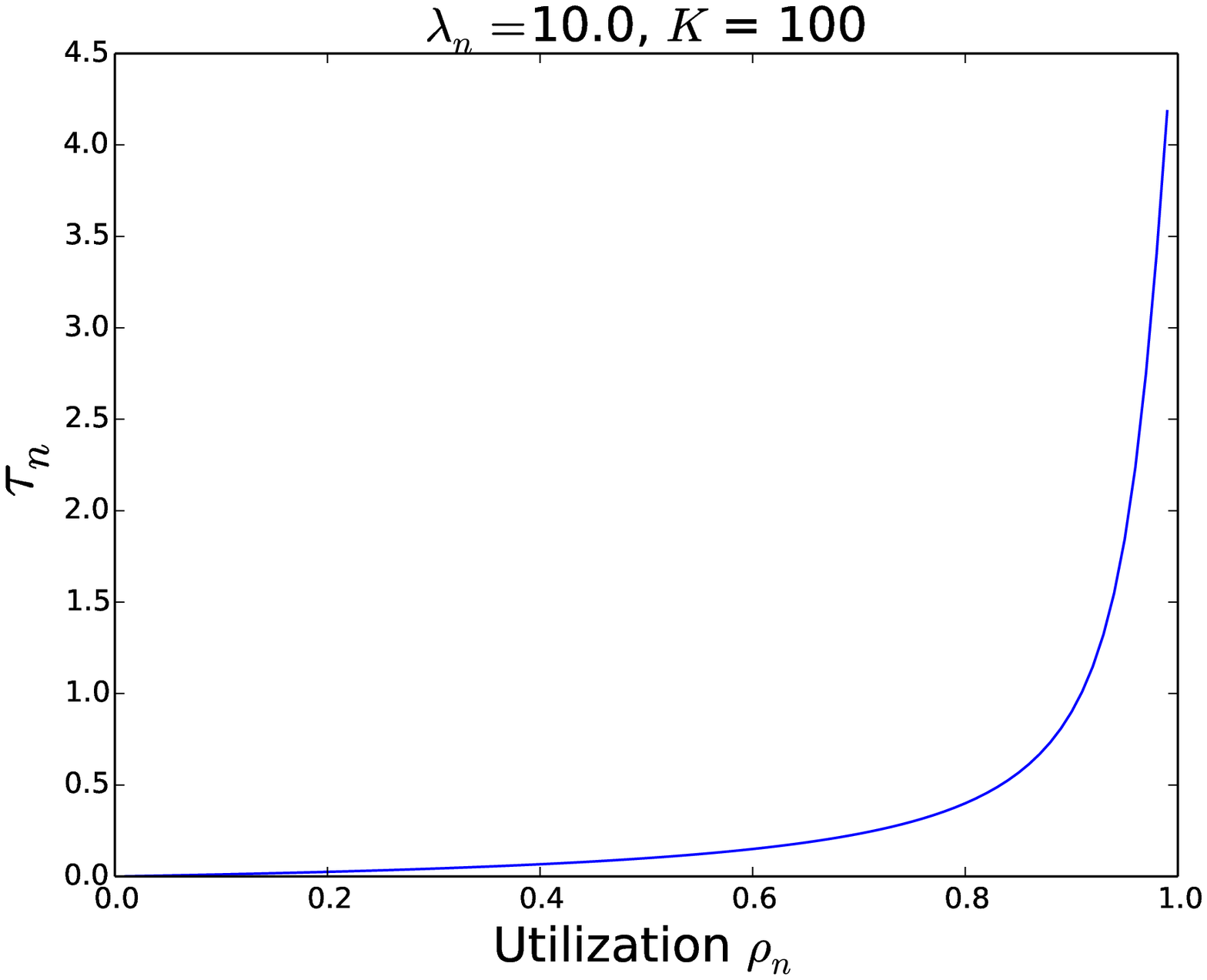}
		\label{fig:tau_rho}
	}
	\subfloat[Packet dropping probability vs. utilization at node $n$.] {
		\includegraphics[width=2.6in]{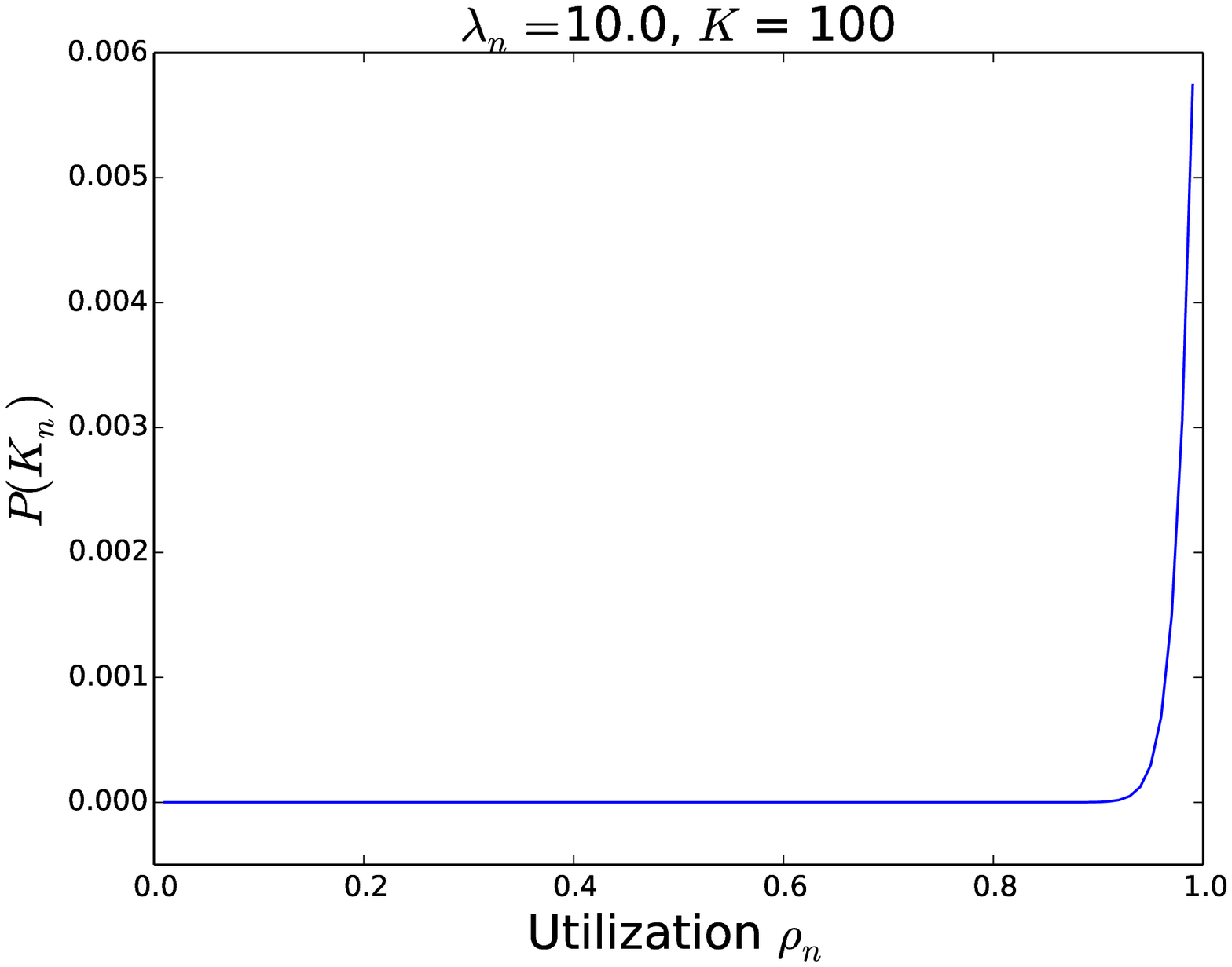}
		\label{fig:drop_rho}
	}
	\caption{Effects of node utilization. From the two figures shown here, one can see that the node utilization $\rho_n$ has a non-linear effect on both the latency and the probability of dropped packets at each node $n$. Beyond a certain threshold, the value of latency and probability grows exponentially. Prior to the placement of VNFs, it is hard to predict where the threshold is. So the observation of these charts motivates our optimization strategy to minimize $\rho_n$ as much as possible across every node $n \in N$ in the network.}
	\label{fig:rho_effects}
\end{figure}

The model to estimate $E(T_c)$ has two key considerations: 1) the latency \red{$\tau_n$} at each node $n \in N_c$, which is independent of the latency at other nodes but is dependent on all service chains' traffic through node $n$ and 2) the probability that a packet may drop at any node $n$, which would require a resend of the packet from the source up to $n$. \red{The retransmission of packets is due to the Transmission Control Protocol (TCP). TCP ensures that all packets will arrive at the destination. If any packet is dropped during transmission, TCP will resend the packet from the source until they reach the destination. The expected latency computation must factor in a packet's expected queueing delay at each node as well as extra time incurred due to resent packets.}

\begin{align}
	E(T_{1 \rightarrow 1}) &= \tau_1\\
	\label{eqn:recursive}
  E(T_{1 \rightarrow n}) &= \tau_{n} + E(R_n) E(T_{1 \rightarrow n - 1}) \textrm{ for } n = 2,\ldots,|N_c|
\end{align}
The recursion in Equation \ref{eqn:recursive} is due to the TCP protocol which resends dropped packets when buffers are full and $E(R_n)$ is the \red{expected} number of retries.
\begin{gather}
	\label{eqn:tau}
	\tau_n = \frac{\rho_n - [1 + K_n (1 - \rho_n)] \rho_n^{K_n + 1}}{\lambda_n (1 - \rho_n) (1 - \rho_n^{K_n})} \\
	\rho_n = \frac{\lambda_n}{\mu_n}
\end{gather}
where $\lambda_n$ is the incoming rate of packets to the node $n$, $\mu_n$ is the rate which the node $n$ processes packets and $K_n$ is the \red{buffer} capacity at node $n$. Equation \ref{eqn:tau} is a standard formula for the M/M/1/K queueing model. Figure \ref{fig:tau_rho} shows $\tau_n$ vs $\rho_n$.
\begin{align}
	E(R_n) &= \frac{1}{1 - P(K_{n})} \\
	P(K_n) &= \frac{1 - \rho_n}{1 - \rho_n^{K_n + 1}} \rho_n^{K_n}
\end{align}
$P(K_n)$ gives the probability that the buffer at node $n$ is full when a packet arrives at node $n$. Appendix \ref{apdx:latency} provides more details on the derivations of these equations and their significance. Figure \ref{fig:drop_rho} shows $P(K_n)$ vs $\rho_n$.

\begin{figure}[htb]
	\centering
	\subfloat[Maximum node utilization vs. number of servers used] {
		\includegraphics[width=3.0in]{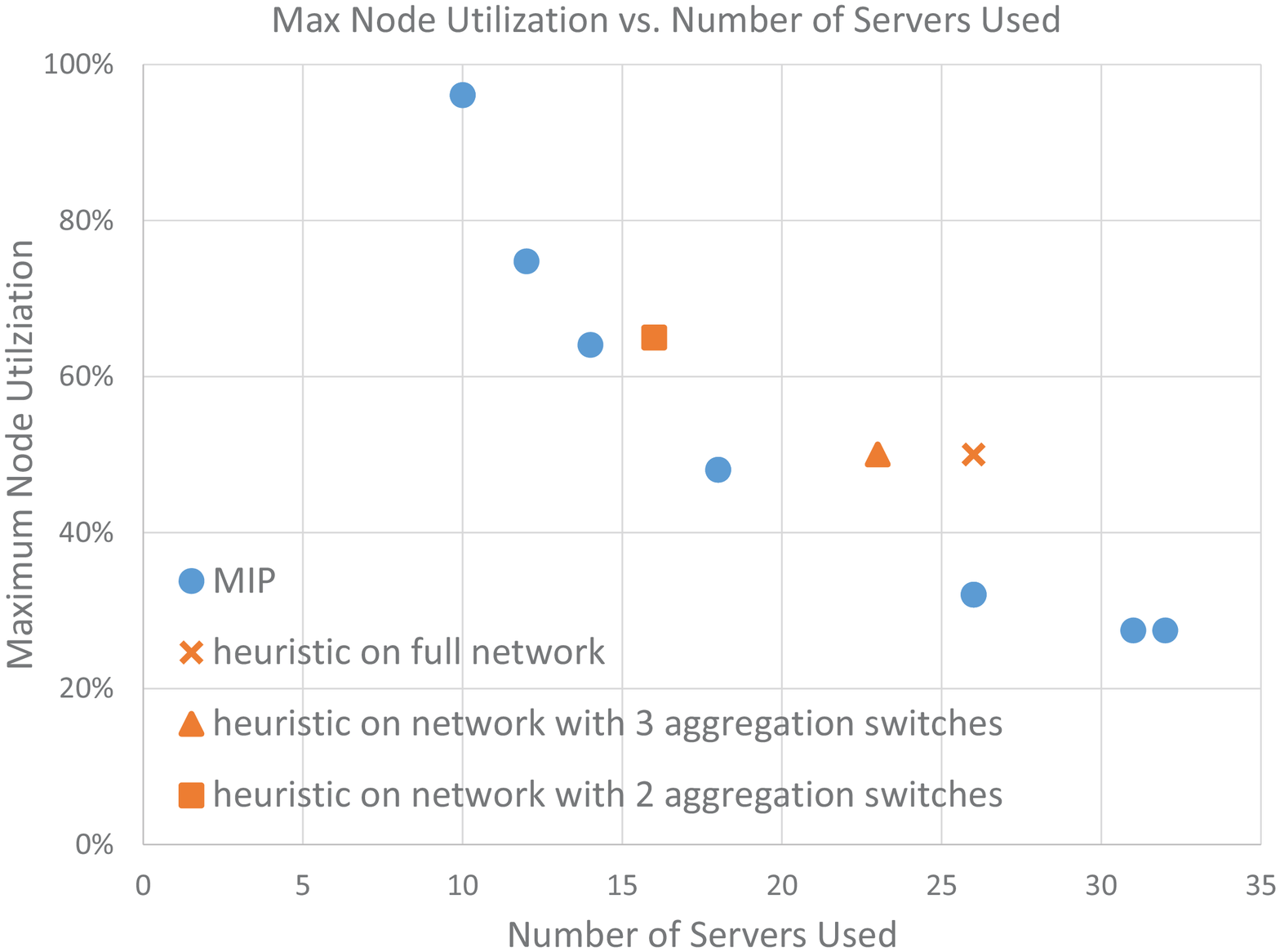}
		\label{fig:efficient_frontier_rho_vs_nodes}
   } \\
   \subfloat[Expected latency vs. number of servers used] {
		\includegraphics[width=3.0in]{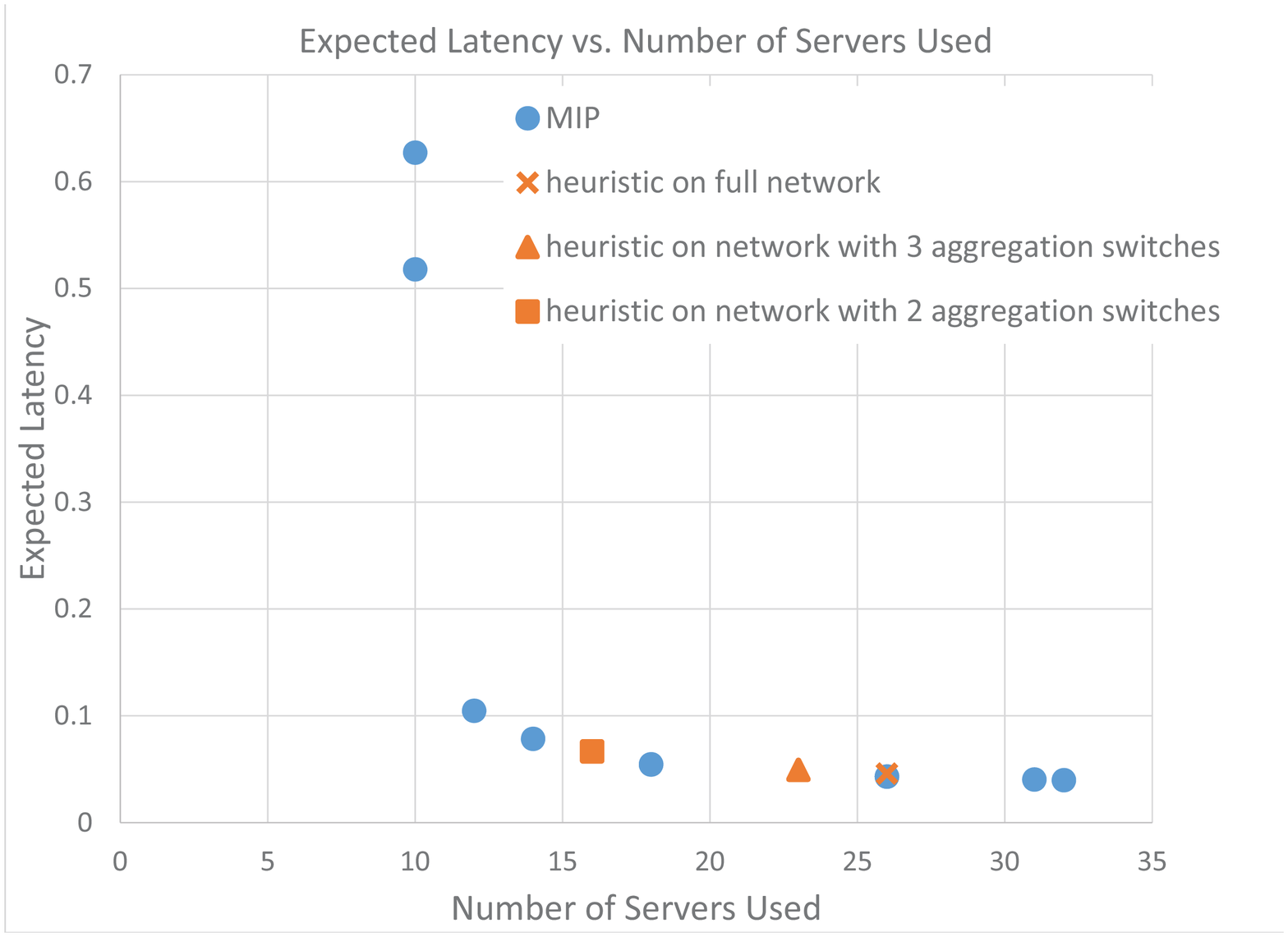}
		\label{fig:efficient_frontier_latency_vs_nodes}
   } \\
   \subfloat[Expected latency vs. maximum node utilization] {
		\includegraphics[width=3.0in]{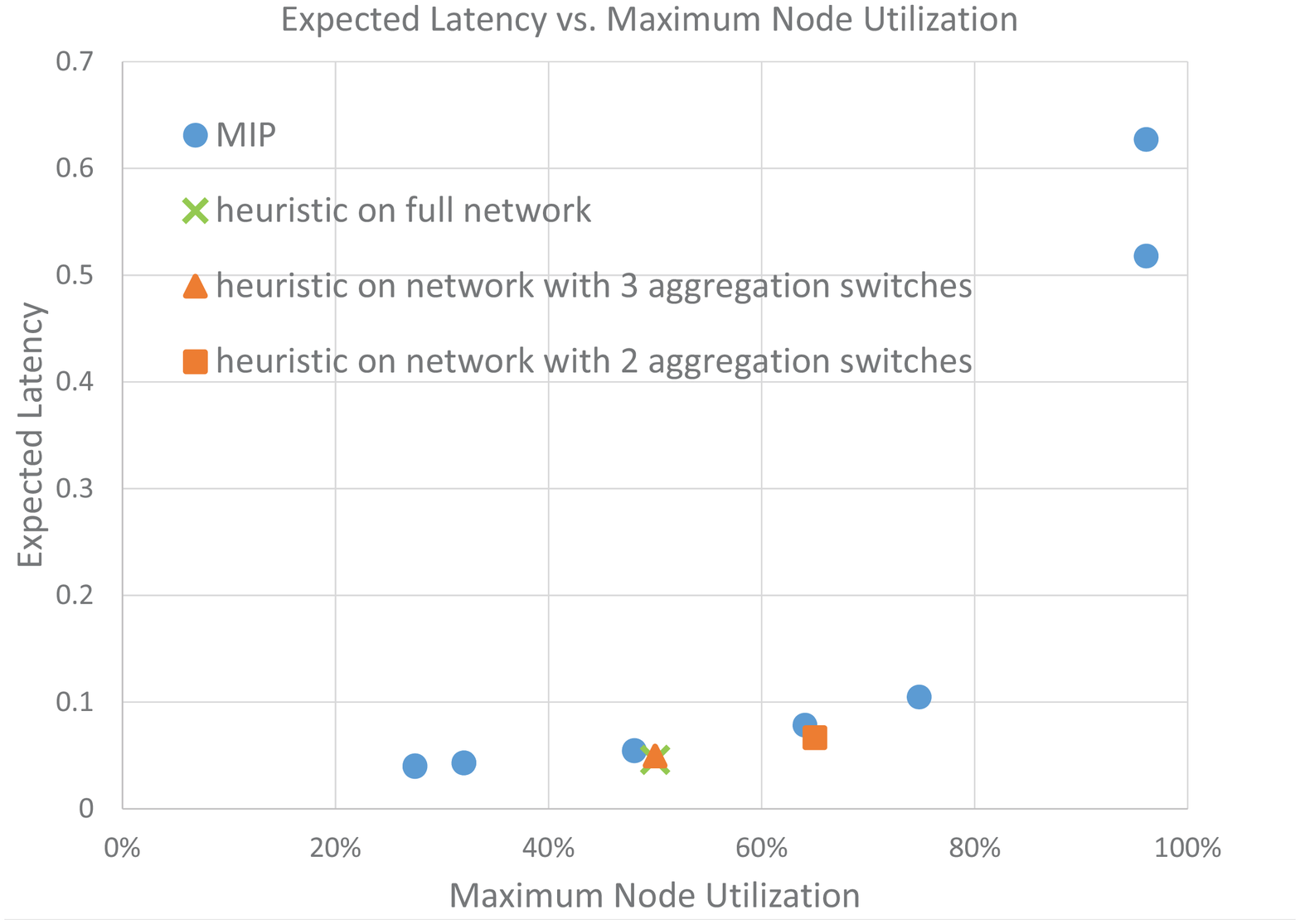}
		\label{fig:efficient_frontier_latency_vs_rho}
   }
  	\caption{Example Solution Metrics}
	\label{fig:efficient frontiers}
\end{figure}

\section{Round Robin Heuristic}
\label{sec:heuristic}

We propose a heuristic that \red{seeks} feasible placements for the incoming service chains while making an effort to minimize the overall latency.
% A flow diagram of \red{the heuristic} is shown in Figure \ref{fig:heuristic_flowchart}.
The basic principle of this heuristic is to distribute the network traffic among different top-of-the-rack (TOR) switches as much as possible, to reduce the maximum node utilization over all nodes. \red{The heuristic achieves this in two ways: first by distributing service chains across TORs and then by limiting the utilization of each machine.}

%\red{To distribute across TORs, the heuristic considers each service chain in succession, by placing consecutive service chains in TOR which is different from the TOR of previous service chain. Each VNF of a service chain is hosted on a machine for the chosen TOR. If the TOR cannot handle all the VNFs for a service chain, then the load is placed on the next TOR.}

\red{Throughout the execution, we maintain an upper limit on machine utilization. The heuristic considers each service chain in succession, placing the chain in a TOR which is different from the TOR of the previous service chain. Each VNF of the service chain is placed on a machine that can accommodate its traffic in the chosen TOR. If a machine has hit its utilization limit, we allow multiple machines that host the same type of VNF to fulfill the service chain. If the machines within a TOR cannot handle all the VNFs for a service chain, then the remaining VNFs are placed on the next TOR. If all machines in the datacenter have been used, the upper limit is adjusted upwards to accommodate more traffic and fulfill more service chains.}

%\red{If a machine has hit its utilization limit, we allow multiple machines that host the same type of VNF to fulfill the service chain. If all the machines in the datacenter have been used, the upper limit will be adjusted upwards so that the datacenter can accomodate more traffic to fulfill more service chains.}

 % For each VNF in a service chain, it looks for a TOR with a machine that can host the VNF, where the TOR is different from previously used TOR. If a single machine cannot handle all of the traffic for the VNF, we allow multiple machines that host the same type of VNF to fulfill the service chain.}

\section{Optimization Method}
\label{sec:mip}

The MIP-based optimization method produces a set of solutions to the SFC provisioning problem, each representing a different tradeoff between network performance and resource usage. Here we provide an overview of the formulation; details are presented in \cite{stringer2016}.

The decisions variables include: binary variables indicating whether an instance of a particular VNF is hosted on a particular server; continuous variables representing the fraction of a service chain's traffic that passes through a particular server and between any pair of servers; continuous variables representing the total bandwidth entering any node in the network; and lastly, the maximum utilization over all nodes in the network. The constraints ensure that flow for each service chain is conserved at each node and that the solution does not use more than the available network resources.

The objective is to minimize a weighted combination of the fraction of servers used to host VNFs and the maximum utilization over all nodes in the network. A weighting parameter $\beta \in [0,1]$ is used to set the relative priority of these two objectives. When $\beta = 0$, the objective reduces to minimizing the maximum utilization over all nodes in the network, thus distributing the traffic as uniformly as possible in order to reduce the highest utilization over all nodes. When $\beta = 1$, the objective becomes minimizing the total number of nodes used to host VNFs. A placement which minimizes the number of VNFs tends to concentrate traffic in part of the network, leaving other network resources unused. Solving the MIP over a range of $\beta \in [0,1]$ yields a set of solutions along the efficient frontier of maximum node utilization and number of servers, each representing a different tradeoff between performance and server usage. For each new value of $\beta$, the preceding solution can be used as a starting point for the MIP run, speeding its execution. The heuristic solution is used as a starting solution for the first MIP.

\red{Our formulation has two novel features compared to prior MIP approaches to SFC provisioning. One is the use of a weighted objective function to generate alternative solutions trading off performance and resource cost. A second novel feature is a method of modeling maximum utilization using only linear constraints. Node utilization is the ratio of a node's incoming bandwidth to its capacity. Both bandwidth and capacity are functions of decision variables (capacity at a node depends on the type of VNF assigned to it), and thus utilization is naturally nonlinear in decision variables. Nonlinear constraints make MIP models significantly less tractable.  We employ a novel approach to linearize the maximum utilization by including constraints for each possible VNF type assigned to a node, and using penalties to activate only the applicable constraints. Details can be found in \cite{stringer2016}.}

\begin{table*}[htb]
    \small
    \centering
    \caption{Summary of Results}

    \label{tbl:results}
    \begin{tabular}{|r|r|c|c|c|c|c|c|c|c|}
        \hline
            \# Servers & \# SFCs & Avg     & Avg       & Avg     & Avg          & Avg          & Avg          & Avg        & Avg \\
                       &         & Random  & Heuristic & MIP     & Random       & Heuristic    & MIP          & Optimality & Latency \\
                       &         & Time(s) & Time(s)   & Time(s) & Success (\%) & Success (\%) & Success (\%) & Gap & Gap \\
        \hline
        8 & 5 & 0.35 & 0.49 & ~~4.63 & 70\% & ~94\% & ~90\% & ~9\% & 0\% \\
        \hline
        16 & 10 & 0.35 & 0.51 & 299.63 & 73\% & ~96\% & 100\% & 17\% & 8\% \\
        \hline
        32 & 20 & 0.36 & 0.49 & 602.42 & 69\% & 100\% & 100\% & 18\% & 7\% \\
        \hline
        64 & 40 & 0.32 & 0.45 & & 75\% & 100\% & & & \\
        \hline
        128 & 80 & 0.37 & 0.52 & & 74\% & 100\% & & & \\
        \hline
        256 & 160 & 0.39 & 0.54 & & 74\% & 100\% & & & \\
        \hline
        512 & 320 & 0.42 & 0.76 & & 74\% & 100\% & & & \\
        \hline
        1024 & 640 & 0.64 & 1.15 & & 76\% & 100\% & & & \\
        \hline
        2048 & 1280 & 1.41 & 5.02 & & 75\% & 100\% & & & \\
        \hline
        4096 & 2560 & 5.30 & 4.59 & & 75\% & 100\% & & & \\
        \hline
    \end{tabular}
\end{table*}

\section{Numerical Results}
\label{sec:example}

\subsection{Efficient Frontier} We present an example of the efficient frontier that is generated by the MIP along with results from the round robin heuristic. This example corresponds to the network in Figure \ref{fig:topology}, in which each cluster of switches is aggregated into a single switch. There are 10 service chains to be deployed, each with up to 4 VNFs.

% A second, larger example illustrates a richer set of choices along the efficient frontier of latency vs number of servers.
% This second

The MIP generates a range of SFC provisioning solutions using between 10 and 32 servers for this example. Properties of these solutions are shown in Figure \ref{fig:efficient frontiers}. Figure \ref{fig:efficient_frontier_rho_vs_nodes} shows MIP solutions along the efficient frontier of server usage vs. maximum node utilization.

\red{Unlike the MIP, the heuristic generally produces only one solution. However, by applying it to successively smaller subnetworks,  we can generate multiple solutions that tradeoff latency and server usage, just as the MIP does. Three such heuristic solutions are also shown in Figure \ref{fig:efficient_frontier_rho_vs_nodes},} corresponding to three different versions of the original network: (1) the full network shown in Figure \ref{fig:topology}, (2) a subnetwork in which one aggregation switch and its descendants are removed, and (3) a subnetwork in which two aggregation switches and descendants are removed.

The heuristic solutions are not far from the efficient frontier, indicating that it achieves low maximum utilization relative to the number of servers it uses to host VNFs. The chart in Figure \ref{fig:efficient_frontier_latency_vs_nodes} shows the same set of solutions, in this case highlighting the tradeoff between expected latency and server usage. Note that the heuristic compares even more favorably to the MIP solutions, in that its solutions lie very close to the MIP solution curve. Figure \ref{fig:efficient_frontier_latency_vs_rho} shows directly how latency varies with maximum node utilization in the MIP and heuristic solutions. In particular, it shows how expected latency of the MIP solutions increases with maximum utilization, and grows steeply as maximum utilization approaches 100\%, as in Figure \ref{fig:tau_rho}. These properties support the choice of maximum node utilization as a good proxy objective for expected latency.

% \subsection{Timings} Figure \ref{fig:timing_vs_servers} shows an illustration of the comparison in timings between heuristic and MIP. The left y-axis gives the timings of the heuristic while the right y-axis gives the timings of the MIP. The x-axis shows the logarithmic number of available servers in the topology for the consideration of VNF placements. By comparing the trend and gradient of the lines in Figure \ref{fig:timing_vs_servers}, one can see that the running time of the MIP grows rapidly when we increase the size of the topology. The results in Figures \ref{fig:efficient frontiers} and \ref{fig:timing_vs_servers} show that the heuristic saves a large amount of time while producing a sub-optimal solution that matches closely to that of the MIP. This suggests that the heuristic should be used in practical situations where the time required for producing VNF placements is important.

 % By taking into consideration these timings, the sub-optimal solution returned by the heuristic becomes an acceptable choice

%\begin{figure}[htb]
%	\centering
%	\includegraphics[width=3.0in]{eps/timing_vs_servers.eps}
%	\caption{Running time of the Heuristic and MIP}
%	\label{fig:timing_vs_servers}
%\end{figure}

% outline

\subsection{Comparison of Solution Quality}
\label{sec:results}

Table \ref{tbl:results} summarizes the timing and results for the round-robin heuristic and the MIP for 100 randomly generated test problems. \red{We also share results for a random placement approach as a baseline.}

We randomly generated 10 problems for each of 10 topologies that vary in number of servers, from 8 through 4096, and number of service chains to be routed. The 10 problems for each topology vary in the specific VNFs and volume of traffic required for each service chain. For each problem, we ran the heuristic, and then ran the MIP with the objective of minimizing the maximum node utilization subject to the constraint that it uses no more servers than the heuristic used for the same problem. This approach allows us to compare the maximum utilization of the heuristic with that of the MIP for a fixed number of servers. \red{We also ran random placement for each problem.}

The average time required by the heuristic is \red{at most 5 seconds}
%less than 1 second
for all topologies, even for problems with over 4,000 servers and 2,000 service chains. For the MIP, we set a time limit of 300 seconds for problems with 16 or fewer servers and 600 seconds for 32-server problems; it is too computationally intensive to run for larger topologies.

Three aspects of solution quality are shown in the table. A first measure is the percentage of service chains deployed. The heuristic may not deploy all service chains, if it runs out of servers. Its success rate, shown in column 5 of Table \ref{tbl:results}, depends on the network capacity, service chain VNF requirements and the service chain traffic requirements. For the MIP we require that all service chains are routed and thus if a solution is found for a given test problem, the MIP service chain deployment rate is 100\%, and otherwise 0\%. \red{Random placement never successfully deployed all service chains; its average success rate was at most 75\%.}

A second measure of quality is in the optimality gap, available only for the set of problems for which the MIP was run. This is the percentage difference between the maximum node utilization achieved by the heuristic and that achieved by the MIP, for the same number of servers used. \red{We present the optimality gap for problems which the heuristic deploys all service chains. The average optimality gap is 18\% or less. The worst case optimality gap over all test problems (not shown in the table) is 34\%.}  \red{Since random placement never deployed all service chains, its solution is not comparable to other approaches that did; we do not include its optimality gap in the table.}
%Note, however, that the heuristic solution deployed fewer service chains than the MIP for some problems.

% We could add the following about latency
A third quality measure is in the latency gap shown for problems on which \red{the MIP and heuristic deployed all service chains. The latency gap} is the percentage difference between the latency of the heuristic solution and the MIP solution as a percentage of MIP latency. The average latency gap %is larger than the optimality gap, reaching up to292\% for the 16-server problems.
% These results underscore the fact that a small increase in node utilization can lead to a large increase in latency, as illustrated in \ref{fig:efficient_frontier_latency_vs_rho}.
\red{ranges from 0-8\%, highlighting the effectiveness of the heuristic and the choice of minimizing maximum utilization as a proxy objective for minimizing latency.}

% Both the heuristic and MIP were implemented in Julia, with Gurobi used as a solver for the MIP.

\section{Related Work}
\label{sec:related}
\textbf{VM placement:} The VM placement has been studied extensively in the cloud computing literature \cite{Lee2014,Ballani:2011:TPD,Jeyakumar:2013:EPN,Jiang:infocom12,Popa:2012:FSN,Xie:2012:OCC}. These work develop heuristics to assign VMs of a tenant close to each other to reduce the overall bandwidth consumption. Different from existing work, our work focus on the placement of VNFs, in the context of service chaining.
% Oktopus~\cite{Ballani:2011:TPD} and FairCloud~\cite{Popa:2012:FSN} focus on providing traffic isolation and optimizing the performance of intra-tenant communications. For dynamic VM placement, ~\cite{Jiang:infocom12} proposes a joint optimization for routing and placement that is light-weight and requires VM small migration cost.

%Another closely related problem is the server replica placement problem, which has been studied by either using simulations~\cite{qiu:infocom01:serverplace,jamin:infocom01:constrained} or network measurements~\cite{Oppenheimer:usenixTech06:wideplace}.
%These work focuses on finding the best locations in the wide area networks for physical web servers. However, they cannot be directly applied to data center and virtual appliances.

\textbf{Service chaining:}
% Service chaining is the process of steering a traffic flow across a predefined set of middleboxes.
Simple \cite{Qazi:2013:SMP} proposes a SDN framework to route traffic through a flexible set of service chains while balancing the load across Network Functions. FlowTags \cite{Fayazbakhsh:2014:ENP} can support dynamic service chaining. Our work is complimentary to these service chain implementation mechanisms. While these work focus on the techniques to realize flexible routing, we provide algorithms that decide on the routes.

\textbf{NFV deployment:}
% NFV allows a VNF to be deployed when and where needed.
% Dynamic NFV routing was by \cite{Gushchin:2015:SRS}.
Lukovszki et al. ~\cite{Lukovszki:2016:MNI} presents an approximate algorithm and an integer programming exact solution. \cite{mohammadkhan2015virtual} also present a MIP model for NFV placement that minimizes the maximum utilization over all links and switches. However, these models considers only the utilization of links and of servers, excluding the switches' utilization. \cite{rajan2015,STEERING2013} measured bottlenecks and maximize throughput in inter-datacenter networks while we focus on intra-datacenter networks. 
% Moreover, they define server utilization not in terms of bandwidth but as the ratio of the number of flows using a server to the number of flows the server can support. They do not consider switch utilization.

\section{Conclusions}
\label{sec:conclusion}
This work addresses the problem of choosing the physical locations of virtual network functions required for service chains, and routing service chain traffic which we term as SFC provisioning problem. We offer a system Stringer consisting of a scalable placement heuristic, an optimization-based approach for generating a series of alternative placement solutions reflecting different tradeoffs between performance and resource usage, and a queueing-theory-based method for estimating average latency per packet under a given VNF placement and routing solution.
%The estimate highlights the fact that minimizing the maximum utilization at switches and servers in the network is a good proxy for minimizing average latency. A second element is a fast, scalable round-robin heuristic that seeks to minimize the maximum node utilization. The third and final element is a MIP-based approach to minimize a weighted combination of number of servers used and maximum node utilization. The MIP approach generates a series of solutions on the efficient frontier of number of servers used and maximum node utilization, allowing a network manager to choose the solution which best reflects his priorities.

 Our experiments comparing the performance of MIP and heuristic show that the heuristic is significantly faster and \red{has an average optimality gap of at most 18\%}. The heuristic can also be used to generate an efficient frontier of solutions, by running it on a succession of subnetworks.

 Stringer has several potential extensions. One way to improve scalability of the optimization is to apply a hierarchical approach, in which we first assign service chains to subnetworks associated with aggregation switches, and then solve the SFC provisioning problem within each subnetwork.  We are also planning to collect data from real networks to validate our queueing theoretic latency estimation.

\bibliographystyle{IEEEtran}
\bibliography{stringer}

\clearpage

\appendix

\section{Additional Details for Expected Latency Derivation}
\label{apdx:latency}
Recall from Section \ref{sec:latency} that we showed the steps to calculate $E(\mathscr{T})$.
\begin{align}
	\label{eq:expected_latency}
	E(\mathscr{T}) &= \sum_{c \in C} \frac {\lambda_c}{\Lambda} E(T_c)
\end{align}
The question now is how to estimate $E(T_c)$, the average amount of time each packet takes to go through the set of services required by service chain $c$. This latency depends on the placement of the service chain's VNFs in the network, the nodes along the paths between successive VNFs, and the amount of traffic (for all service chains) through each of those nodes.   Because the VNF placement and chain routing has already been determined, we know which nodes (switches and servers) $N_c$ that packets of $c$ will flow through. Denote the sequence of nodes in $N_c$ as $\{1,2,\ldots,n,\ldots,|N_c|\}$ where $n$ indicates the $n$th node that the packets will pass through, $1$ as the source of the packets and $|N_c|$ as the final destination of the packets.

The model to estimate $E(T_c)$ has two key considerations: 1) the latency at each node $n \in N_c$, which is independent of the latency at other nodes but is dependent on all service chains' traffic through node $n$ and 2) the probability that a packet may drop at any node $n$, which would require a resend of the packet from the source up to $n$.

\paragraph*{Queueing Theoretic Latency Estimation} We model each node $n$ as a finite capacity, single server queue where packets are processed one at a time while other packets wait in the queue of size $K_n - 1$ for their turn to be processed on a First-In-First-Out (FIFO) policy. We assume that the packets arrive at node $n$ according to a Poisson process with rate $\lambda_n$, a rate which reflects the traffic from all service chains routed through node $n$.

Unlike the case of M/M/1 queues, we model each node with a finite queue size, and packets can drop from the node if the queue is full. Such packet drops result in an outgoing rate that is less than the incoming rate. Although it is possible to derive an approximation for the outgoing rate for an acyclic network, typical networks in data centers have cyclic dependencies between the outgoing and incoming rate, which makes it hard to estimate it correctly analytically. An example of a network cyclic dependency is traffic which flow through a switch and would later flow back to the same switch after processing at the leaves beneath the switch itself. To simplify the model, we assume that the outgoing rate from node $n$ is equal to the incoming rate.

% \textcolor{red}{Somewhere we need to justify this assumption, since having a network of M/M/1/K queues may not guarantee this.}
The time to process each packet at the node $n$ (excluding queueing time) follows an exponential distribution $\mu_n$, and it is assumed that $\lambda_n < \mu_n$. These assumptions allow us to use the well-known formula in the M/M/1/K queueing literature to estimate the expected latency $\tau_n$ of a packet at node $n$, including both queueing time and service time at $n$.
\begin{gather}
	\label{eq:mm1k_T}
	\tau_n = \frac{\rho_n - [1 + K_n (1 - \rho_n)] \rho_n^{K_n + 1}}{\lambda_n (1 - \rho_n) (1 - \rho_n^{K_n})} \\
	\rho_n = \frac{\lambda_n}{\mu_n}
\end{gather}

\paragraph*{Packet Loss at Each Node $n$} Since we model each node $n$ with a finite capacity queue of length $K_n - 1$, packets that arrive to find the queue full will be discarded. The probability of packets dropping in this manner is equal to the probability that there are $K_n$ packets in the system (one packet being processed and $K_n - 1$ packets in the queue). For an M/M/1/K queue, the probability of having $K_n$ packets in the system is,
\begin{gather}
	\label{eq:pK}
	P(K_n) = \frac{1 - \rho_n}{1 - \rho_n^{K_n + 1}} \rho_n^{K_n}
\end{gather}

In software applications that use the Transmission Control Protocol (TCP) for transferring network packets, TCP ensures that all packets will arrive at the destination. If any packet is dropped during transmission, TCP will resend the packet from the source until they reach the destination. The expected latency computation must factor in a packet's expected queueing delay at each node as well as extra time incurred due to resent packets. Let $E(T_{1 \rightarrow n})$ represent the expected latency for a packet to visit the sequence of nodes as $\{1,2,\ldots,n\}$ in $N_c$, for $n = 1,2,\ldots,|N_c|$. Thus, $E(T_c) = E(T_{1 \rightarrow |N_c|})$.  We define a recursive formula for the latency as follows:
\begin{align}
	E(T_{1 \rightarrow 1}) &= \tau_1\\
    E(T_{1 \rightarrow n}) &= \tau_{n} + E(R_n) E(T_{1 \rightarrow n - 1}) \textrm{ for } n = 2,\ldots,|N_c|
    \label{eq:recursion}
\end{align}
where $E(R_n)$ is the expected number of resends required to transmit the packet from node $1$ to node $n$. To compute $E(R_n)$, note that $P(R_n = m)$ is the probability of the packet dropping $m-1$ times at node $n$ and succeeding on the $m$th time. Thus $E(R_n)$ is derived as follows:
\begin{align}
	P(R_n = m) &= P(K_{n})^{m-1} \left[ 1-P(K_{n}) \right] \\
	E(R_n) &= \sum_{m=1}^{\infty} m \cdot P(R_n = m) \\
	\label{eq:Rn}
	&= \frac{1}{1 - P(K_{n})}
\end{align}
Using Equations \ref{eq:expected_latency} to \ref{eq:Rn}, we can evaluate the expected latency $E(\mathscr{T})$.

While the derivation in this section allows us to evaluate the expected latency of each service chain given a particular SFC provisioning solution, it does not lend itself to optimizing for latency when making placement decisions, since the placement decisions (and implied congestion) affect latency in a complex and nonlinear way.

% \begin{figure*}[htb]
% 	\centering
% 	\subfloat[Latency vs. utilization at node $n$.] {
% 		\includegraphics[width=2.6in]{eps/exp_latency_vs_rho_with_K.eps}
% 		\label{fig:tau_rho}
% 	}
% 	\subfloat[Packet dropping probability vs. utilization at node $n$.] {
% 		\includegraphics[width=2.6in]{eps/prob_drop_vs_rho_with_k.eps}
% 		\label{fig:drop_rho}
% 	}
% 	\caption{Effects of node utilization. From the two figures shown here, one can see that the node utilization $\rho_n$ has a non-linear effect on both the latency and the probability of dropped packets at each node $n$. Beyond a certain threshold, the value of latency and probability grows exponentially. Prior to the placement of VNFs, it is hard to predict where the threshold is. So the observation of these charts motivates our optimization strategy to minimize $\rho_n$ as much as possible across every node $n \in N$ in the network.}
% 	\label{fig:rho_effects}
% \end{figure*}

However, the derivation offers insight into the importance of node utilization in expected latency.  Consider, for example, the relationship \label{eq:mm1k_T} between the expected latency $\tau_n$ at node $n$ and the utilization $\rho_n$ at node $n$, illustrated in Figure \ref{fig:tau_rho} for arrival rate $\lambda_n = 10$ and queue capacity $K_n = 100$.  Latency grows steeply as utilization approaches $100\%$. Moreover, the relationship \label{eq:pK} between packet dropping probability $P(K_n)$ and node utilization $\rho_n$ reveals the importance of utilization $\rho_n$ in preventing packets from being dropped. Figure \ref{fig:drop_rho} illustrates how the packet dropping probability $P(K_n)$ grows abruptly with as node utilization approaches $100\%$ under the same assumptions on $\lambda_n$ and $K_n$.

These objectives suggest a simple but powerful objective to use in SFC provisioning. By making placement decisions to minimize the maximum node utilization in the physical network, we can both avoid packet loss and reduce latency. The SFC provisioning methods described below pursue the goal of minimizing the maximum node utilization.

%[Mixed Integer Program Constraints]
\section{Additional Details for Mixed Integer Program}
% \section{Additional Details for Expected Latency Derivation}
% \label{apdx:latency}

% \section{Additional Details for Mixed Integer Program}
\label{apdx:mip}

This appendix describes the contraints of the MIP formulation introduced in Section \ref{sec:mip}.

\textbf{Model Parameters:}

\begin{itemize}
  \item $N$: the set of all nodes in the network (servers and switches).
  \item $L \subset N$: the set of servers, which are leaves in the tree network.
  \item $r \in N$: the root node.
  \item $P_{n,m}$: the set of nodes in the unique acyclic path from node $n$ to $m$, including the destination $m$ but excluding the origin $n$.
  \item $\mu_n$: the processing rate, in packets per second, associated with switch $n \in N \setminus L$.
  \item $S$: the set of different server types.
  \item $s_l \in S$: the machine type associated with server $l \in L$.
  \item $V$: the set of VNF types. Instances of these VNF types must to be assigned to servers in the physical network in order to accommodate service chains.
  \item $\gamma^s_v$: the processing rate, in packets per second, of VNF type $v \in V$ when assigned to server type $s \in S$.
  \item $C$: the set of service chains to be mapped to the network.
  \item $q_c$: the length of the sequence of VNFs in service chain $c$.
  \item $\alpha^c_{i,v}$: a binary parameter indicating whether the $i$th service in chain $c$ is of type $v$.
  \item $\lambda_c$: arrival rate, in packets per second, for chain $c$.
  \item $M$: a large positive scalar. For example, any $M > \max \{1,\max_{s,v} \{\gamma^s_v\} \}$ is suitable.
  \item $\beta \in [0,1]$: a parameter representing the relative weight between two metrics, number of servers used and maximum utilization, in the objective function.
\end{itemize}

\textbf{Decision Variables:} The decision variables describe the assignment of VNF instances to leaf nodes, the mapping of each service chain to one or more paths in the network, the volume of flow for each chain along each of its paths, the rate of traffic into each node, and performance metrics associated with the solution.

\begin{itemize}
    %\item $d_c \in \{0,1\}$ indicates whether service chain $c$ is deployed in the physical network. All service chains will be deployed if it is possible to do so.
    \item $x_{v,l} \in \{0,1\}$ indicates whether an instance of VNF type $v$ is placed on leaf $l$.
    \item $y^c_{i,l} \in [0,1]$ is the fraction of traffic for the $i$th function in service chain $c$ that is served by leaf node $l$.
    \item $z^c_{i,k,l} \in [0,1]$ is the fraction of traffic going from the $i$th to $(i+1)$st function in service chain $c$ that travels from leaf node $k$ to leaf node $l$.
    \item $b_k \ge 0$ is the total traffic rate in packets per second into node $k \in N$.
    \item $\rho$ is the maximum node utilization over all nodes in the network.
\end{itemize}

\paragraph*{Constraints} The MIP constraints ensure that flow for each service chain is conserved at each node, that the solution does not use more than the available network resources, and that the maximum utilization metric is measured.

\begin{align}
    \label{eq:one_vnf_type_per_leaf}
    \sum_{v \in V} x_{v,l} \le 1 &~ , \quad l \in L \\
    \label{eq:need_right_type_of_vnf}
    y^c_{i,l} \le \sum_v \alpha^c_{i,v} ~ x_{v,l} &~ , \quad c \in C, i \le q_c, l \in L    \\
    \label{eq:each_chain_vnf_must_be_placed}
    \sum_{l \in L} y^c_{i,l} = 1 &~ , \quad c \in C, i \le q_c \\
    \label{eq:z_at_most_first_y}
    z^c_{i,k,l} \le y^c_{i,k} &~ , \quad c \in C, i < q_c, k,l \in L \\
    \label{eq:z_at_most_second_y}
    z^c_{i,k,l} \le y^c_{i+1,l} &~ , \quad c \in C, i < q_c, k,l \in L \\
    \label{eq:total_outflow_allocated}
    \sum_{k,l \in L} z^c_{i,k,l} =  1 &~ , \quad c \in C, i < q_c
\end{align}
\begin{gather}
    \label{eq:flow_conservation_at_leaves}
    y^c_{1,k} + \sum_{\substack{m \in L \\ i < q_c}} z^c_{i,m,k} = \sum_{\substack{m \in L \\ i < q_c}} z^c_{i,k,m} + y^c_{q_c,k} ~ , \quad c \in C, k \in L \\
    b_k = \sum_{c \in C} \lambda_c \left(1 + \sum_{\substack{l \in L:\\ k \in P_{r,l}}} y^c_{1,l} + \sum_{\substack{m \in L: \\k \in P_{m,r}}} y^c_{q_c,m} + \sum_{\substack{i < q_c\\l,m \in L: \\k \in P_{l,m}}}  z^c_{i,l,m} \right) ~ , \nonumber \\
    \label{eq:bandwidth_into_node}
    \quad k \in N \setminus {r} \\
    \label{eq:bandwidth_into_root}
    b_r = \sum_{c\in C} \lambda_c \left(1 + \sum_{m \in L} y^c_{q_c,m} + \sum_{\substack{i < q_c\\l,m \in L:\\ r \in P_{l,m}}} z^c_{i,l,m}\right)
\end{gather}
\begin{align}
    \label{eq:lambda_less_than_switch_mu}
    b_n \le \mu_n &~ , \quad n \in N \setminus L  \\
    \label{eq:lambda_less_than_vnf_mu}
    b_l \le \sum_{v \in V} \gamma^{s_l}_v ~ x_{v,l} &~ , \quad l \in L \\
    \label{eq:max_utilization_switch}
    \rho \ge \frac{b_n}{\mu_n} &~ , \quad n \in N \setminus L \\
    \label{eq:max_utilization_leaf}
    \rho \ge \frac{b_l}{\gamma^{s_l}_v} - M(1-x_{v,l}) &~ , \quad l \in L, v \in V
\end{align}

The constraint \ref{eq:one_vnf_type_per_leaf} ensures that each server $l \in L$ can have at most one VNF type assigned to it. Constraint \ref{eq:need_right_type_of_vnf} enforces that the $i$th function in service chain $c$ can only be placed on a server hosting the VNF type associated with the $i$th function. Constraint \ref{eq:each_chain_vnf_must_be_placed} requires that then the total traffic for its $i$th function must be placed.

We need inequalities \ref{eq:z_at_most_first_y} and \ref{eq:z_at_most_second_y} to ensure that  $z^c_{i,k,l}$ does not exceed $y^c_{i,k}$ or $y^c_{i+1,l}$  for each chain $c$, for each function index $i < q_c$, and each physical server pair $k,l \in L$. Constraint \ref{eq:total_outflow_allocated} implies that the total required traffic rate from the $i$th function to the $(i+1)$st function in service chain $c$ must be allocated. Flow conservation constraint \ref{eq:flow_conservation_at_leaves} requires that the traffic for service chain $c$ into server $k$ (the left hand side) must equal the traffic service chain $c$ exiting $k$. Constraint \ref{eq:bandwidth_into_node} defines the total traffic rate $b_k$ into each non-root node $k \in N \setminus \{r\}$. For a given service chain $c$, the first term ($\lambda_c \sum_{l \in L: k \in P_{r,l}} f^c_l$) captures the traffic into switch $k$ coming from the root to any server $l$ hosting the first function in the chain, the second term ($\lambda_c  \sum_{m \in L: k \in P_{m,r}} h^c_m$) captures traffic into switch $k$ heading toward the root from any server $m$ hosting the last function in the chain, and the remaining term captures traffic between any pair of servers $l$ and $m$ hosting consecutive functions in the chain for which their path passes through switch $k$. Constraint \ref{eq:bandwidth_into_root} defines the total traffic rate $b_r$  into the root node $r$. The first term captures the traffic into the root $r$ coming from outside the network ($\lambda_c$), the second describes the traffic into $r$ from any server $m$ hosting the last function in the chain (term $\lambda_c \sum_{m \in L} y^c_{q_c,m}$), and the final term captures traffic between any pair of servers $l$ and $m$ hosting consecutive functions in the chain for which their path passes through the root.

Constraints \ref{eq:lambda_less_than_switch_mu} and \ref{eq:lambda_less_than_vnf_mu} enforce that the traffic rate into a switch or a server must not exceed the available processing rate. In the case of inequality \ref{eq:lambda_less_than_vnf_mu}, the server's processing rate is governed by the VNF assigned to it. Constraints \ref{eq:max_utilization_switch} and \ref{eq:max_utilization_leaf} help define the maximum utilization $\rho$ over network resources: $\rho$ must be at least as great as the utilization at any switch $n \in N \setminus L$, and at least as great as the utilization at any server $l \in L$. Because the processing rate of a server $l$ depends on the VNF $v$ assigned to it, we must have a separate constraint of type \ref{eq:max_utilization_leaf}  for $l \in L$ and $v\in V$. If VNF type $v$ is assigned to server $l$, then $M(1-x_{v,l}) = 0$ and \ref{eq:max_utilization_leaf} requires that $\rho \ge b_l / \gamma^{s_l}_v$, where $\gamma^{s_l}_v$ is the processing rate of VNF $v$ if assigned to the server $l$. If $v$ is not assigned to server $l$, then the right hand side of \ref{eq:max_utilization_leaf} is negative, and so imposes no restriction on $\rho$.

Note that while no constraint forces $\rho$ to equal the maximum utilization over all network nodes, the objective function will drive the value of $\rho$ down to the smallest value satisfying the constraints \ref{eq:max_utilization_leaf} and \ref{eq:max_utilization_switch}, thus ensuring that it equals the true maximum utilization over all nodes in the network.

\textbf{Model Objectives:} The objective is to minimize a weighted combination of the number of nodes utilized and the maximum utilization over all nodes in the network.

\begin{equation}
    \label{eq:weighted_objective}
    w = (1 - \beta) \rho + \beta \frac{1}{|L|} \sum_{v \in V, l \in L} x_{v,l}
\end{equation}

When $\beta = 0$, the objective reduces to minimizing the maximum utilization over all nodes in the network. This choice of objective has the effect of distributing the traffic as uniformly as possible in order to reduce the highest utilization over all nodes. If instead  $\beta = 1$, the objective becomes minimizing the total number of nodes used to host VNFs. A placement which minimizes the number of VNFs tends to concentrate traffic in part of the network, leaving other network resources unused. Solving the MIP over a range of $\beta \in [0,1]$ yields a set of solutions that represent different tradeoffs between performance and server usage.

\paragraph*{Extensions}
There are several possible extensions to the MIP model. One such extension is handling the case that only a subset of the service chains can be deployed. It may happen that not all service chains can be accommodated by the network. In that case, we still want to produce a solution that deploys a subset of service chains. We assume that there is a priority order among service chains. Let $\pi_c$ denote the priority weight of service chain $c$, where higher priority weight corresponds to higher priority. We introduce a new binary decision variable $d_c \in \{0, 1\}$ for each service chain $c \in C$ indicating $c$ is deployed in the physical network. For constraints \ref{eq:each_chain_vnf_must_be_placed} and \ref{eq:total_outflow_allocated}, we change the right hand side to $d_c$. We also introduce a new constraint that ensures service chains are deployed according to the given priority:
\begin{equation}
    \label{eq:service_chain_priority}
    d_c \ge d_{c'}
\end{equation}
for all service chains $c,c' \in C$ for which $\pi_c > \pi_{c'}$.

In this extension, the primary objective is to deploy all service chains if possible, and if not, to deploy as many service chains as possible according to the given priority. To that end, our solution procedure would change slightly. We would first solve the MIP (including the new decision variables and constraints) with the objective of maximizing $w' = \sum_c d_c$, the number of service chains deployed. We then refine the solution by re-solving with the objective in Equation \ref{eq:weighted_objective} while fixing the $d_c$ variables to the values obtained in the first solution.

Other extensions that can be easily accommodated include:
\begin{itemize}
  \item Imposing constraints on link bandwidth.
  \item Limiting the length of the path(s) travelled by a service chain.
  \item Including edge utilization when computing maximum utilization $\rho$.
  \item Allowing multiple VNFs to be hosted on each server.
  \item Requiring that service chain traffic flows are not split across multiple paths. (This extension requires continuous variables to become binary.)
  \item Enforcing redundancy by prohibiting select pairs of service chains from sharing subnetworks.  For example, we could restrict a pair of service chains from using servers under a common TOR switch or aggregation switch.
  \item Deploying additional VNFs and service chains while keeping existing deployments fixed.
\end{itemize}

% or


% \appendices

% \section{}

% \section*{Acknowledgment}



%\vfill


\end{document}